# Deep Learning-Based Photoacoustic Imaging of Vascular Network Through Thick Porous Media

Ya Gao, Wenyi Xu, Yiming Chen, Weiya Xie, and Qian Cheng, *Member, IEEE*

*Abstract*—Photoacoustic imaging is a promising approach used to realize *in vivo* transcranial cerebral vascular imaging. However, the strong attenuation and distortion of the photoacoustic wave caused by the thick porous skull greatly affect the imaging quality. In this study, we developed a convolutional neural network based on U-Net to extract the effective photoacoustic information hidden in the speckle patterns obtained from vascular network images datasets under porous media. Our simulation and experimental results show that the proposed neural network can learn the mapping relationship between the speckle pattern and the target, and extract the photoacoustic signals of the vessels submerged in noise to reconstruct high-quality images of the vessels with a sharp outline and a clean background. Compared with the traditional photoacoustic reconstruction methods, the proposed deep learning-based reconstruction algorithm has a better performance with a lower mean absolute error, higher structural similarity, and higher peak signal-to-noise ratio of reconstructed images. In conclusion, the proposed neural network can effectively extract valid information from highly blurred speckle patterns for the rapid reconstruction of target images, which offers promising applications in transcranial photoacoustic imaging.

## I. Introduction

HUMAN brain mapping is one of the most exciting research areas in contemporary medical imaging. Photoacoustic imaging (PAI) is a promising medical imaging modality that uses a short electromagnetic pulse to generate ultrasound waves based on the thermoelastic effect. The imaging mechanism based on the photoacoustic (PA) effect achieves high contrast in optical imaging and high spatial resolution in ultrasound imaging. The high contrast is attributed to the high sensitivity of biomolecules, such as hemoglobin [1]–[4], to light absorption due to the specificity of energy level differences in biomolecules [5]. Further, compared to the round-trip transmission attenuation in ultrasound imaging, the high spatial resolution is attributed to the only one-way transmission attenuation of ultrasonic waves during the PA emission phase. Wang *et al.* were the first to apply PA tomography to the brain imaging of mice; they successfully obtained the structural images of the mouse brain vessel with a penetration depth of 8 mm and a resolution of 0.2 mm [1]. Tang *et al.* developed a wearable, multispectral, three-dimensional PA brain imaging system for rats with a resolution of 0.2 mm [6]. Subochev *et al.* [7], Yang *et al.* [8], and Li *et al.* [9] also reported outstanding works in the field of PA brain imaging. However, these studies focused on mice and rats with thin skulls (thickness: 0.3–0.8 mm), rather than human skulls. Because the human skull (thickness: 2–11 mm) is the thick porous media that will lead to the strong scattering of PA signals, it is difficult to directly achieve high-resolution PAI of the brain.

Imaging through thick porous media has been a major challenge in the field of medical imaging. Because of the strong scattering of the skull, both light and ultrasound imaging suffer from the interference of multiple scattering times, which leads to the distortion of the propagation direction and wavefront information. Thus, it is difficult to achieve high-resolution images to observe changes in tissue morphology and structure, which are caused by intracranial disease or injury. For photoacoustic brain imaging, the frequency of the photoacoustic signal is closely related to the size of the blood vessel [10], [11], that is, the larger the size distribution range, the wider the frequency of the photoacoustic signal. In this work, we focused on cerebral microvessels with a diameter of 100–200 $\mu$m; therefore, the excited photoacoustic signal has a wide bandwidth in the range of 7.5–15 MHz with the sound speed of soft tissue of 1500 m/s. When transmitted to the skull, their wavelengths are approximately within the range of 167–47 $\mu$m according to the sound speed of the skull (2500–4100 m/s [12], [13]), which is in the same order as the skull pore size [12], [14]–[19] (100–500 $\mu$m in this study), so it is *Mie scattering* here [20], [21]. Unfortunately, unlike Rayleigh scattering, Mie scattering is accompanied by

Manuscript received 17 December 2021; revised 7 February 2022; accepted 6 March 2022. Date of publication 16 March 2022; date of current version 1 August 2022. This work was supported in part by the National Natural Science Foundation of China under Grant 12034015 and Grant 11827808, in part by the Shanghai Municipal Science and Technology Major Project under Grant 2021SHZDZX0100, and in part by the Shanghai Municipal Commission of Science and Technology Project under Grant 19511132101. *(Corresponding author: Qian Cheng.)*
Ya Gao, Wenyi Xu, Yiming Chen, and Weiya Xie are with the Institute of Acoustics, School of Physics Science and Engineering, Tongji University, Shanghai 200092, China, and also with the MOE Key Laboratory of Spine and Spinal Cord Injury Repair and Regeneration, Tongji University, Shanghai 200086, China (e-mail: y.gao@tongji.edu.cn; dolores@tongji.edu.cn; y.m.chen@tongji.edu.cn; weiyaxie@tongji.edu.cn).
Qian Cheng is with the Institute of Acoustics, School of Physics Science and Engineering, Tongji University, Shanghai 200092, China, also with the MOE Key Laboratory of Spine and Spinal Cord Injury Repair and Regeneration, Tongji University, Shanghai 200086, China, and also with the Frontiers Science Center for Intelligent Autonomous Systems and the Shanghai Key Laboratory of Autonomous Intelligent Systems, Tongji University, Shanghai 201210, China (e-mail: q.cheng@tongji.edu.cn).
Digital Object Identifier 10.1109/TMI.2022.3158474

some side lobes on both sides of the main lobe, so the beamforming process is more complicated. Therefore, transcranial photoacoustic microvascular imaging combines the broadband scattering problem in the frequency domain with the main lobe and side lobes signal superimposition problems in the spatial domain, see Appendix A for more details.

Over the years, a variety of algorithms have been developed to solve ultrasound scattering problem in complex media. The time-reversal [22], [23] of the ultrasonic field is a simple and robust method designed to propagate the distorted wavefield through an inhomogeneous medium in the reverse direction. Using the invariance of the wave equation in a lossless medium, the distorted wave field of the source at the desired focal point is recorded through a time-reversal mirror (TRM) and transmitted in the reverse direction to best focus on the source. Based on this principle, a variety of improved algorithms have been proposed, such as array imaging [24], inverse filter [25], and combined amplitude compensation and conventional steering to extend the method to the skull [26]. In addition, Aubry and Derode [27]–[29] separated the single and multiple scattering in the backscattered wave based on the random matrix theory and further applied the time-reversal operator to decompose and eliminate the contribution in addition to the multiple scattering.

Recently, the rise of deep learning provides a new approach to improve the quality of photoacoustic imaging, such as sparse data or limited field of view imaging [30]–[35], artifact removal [36], [37], and limited bandwidth of transducer [38]–[40]. Razansky *et al.* used under-sampling data or photoacoustic images reconstructed from limited-view scans. The trained U-Net can enhance the visibility of arbitrary directional structures and restore the expected image quality [30]. Allman *et al.* used convolutional neural networks (CNNs) to locate and distinguish photoacoustic point sources and artifacts, thereby removing artifacts from the photoacoustic data channel [36]. Awasthi *et al.* proposed a U-Net based architecture for bandwidth enhancement of photoacoustic signals collected at the boundary of the domain [40]. However, the above-mentioned researches are all about acoustic wave propagation in soft tissues rather than bone tissues, because these methods are difficult to be directly transplanted into solid-liquid two-phase bone tissues with a large number of strong scattering interfaces. In addition, deep learning has many breakthroughs in solving the problem of scattering in optical imaging [41]–[43]. Lyu *et al.* built a hybrid neural network to achieve the recovery of handwritten digits and letters through a strong scattering medium [41]. Li *et al.* used trained neural networks to predict untrained scattering media and different types of imaging targets, successfully reconstructing object images [42]. These achievements demonstrate that neural networks can autonomously mine and utilize the hidden physical statistical invariance based on high-quality data. However, there is currently almost no research on using deep learning to solve the problem of sound scattering.

In this study, based on the high sensitivity of PAI to hemoglobin, we carry out a study on transcranial cerebral vascular network photoacoustic imaging, and attempt to solve the problem of strong scattering of human skull to photoacoustic signals. Here, the received transcranial photoacoustic signals used for the reconstruction of cerebrovascular images have the following dual characteristics of the frequency and spatial domains. In the frequency domain, the photoacoustic signals generated from the distributed vascular size are broadband (i.e., the wavelength of photoacoustic signals is distributed); and in the spatial domain, the transcranial signal field, which is scattered and superimposed by the size-distributed holes, has a complex spatial distribution. We innovatively use deep learning algorithms and verify the feasibility of using this method to solve the problem of image restoration by reconstructing photoacoustic signals through the scattering layer. We built a convolutional neural network (CNN) to extract "hidden" effective PA information in the speckle patterns obtained from vascular network under thick porous scattering media. High-quality PA images of vascular network were obtained using the CNN in both simulations and experiments.

The remainder of this paper is organized as follows. Section II analyzes the generation and propagation of transcranial PA signals, including the physical model of PAI reconstruction and how to use CNN to optimize low-quality images. This section also presents the proposed network architecture, simulation, and experimental data acquisition in detail. Section III provides details of the reconstruction results of simulations and experiments, as well as evaluation indicators. Section IV discusses the results and limitations of the proposed method. Finally, Section V summarizes the research and outlines the prospects for future work.

## II. Materials and Methods
### A. Photoacoustic Brain Imaging Model

The physical transmission process of transcranial vascular PAI is shown in Fig. 1. 1) On the outer side of the skull, a pulsed laser emits laser light at the characteristic absorption wavelength of hemoglobin, and the light energy passes through the skull. 2) Transcranial light is scattered by the pores in the skull and it irradiates the brain tissue. 3) The hemoglobin molecules in the intracranial blood vessels are excited by the scattered light to generate vibrations. Then the ultrasound wave is generated based on the PA effect. 4) The ultrasound wave carrying blood vessel information propagates outside the skull and is strongly scattered by the skull. 5) An array transducer outside the skull is used to receive the scattered ultrasound wave for intracranial blood vessel distribution reconstruction.

In step 3), the conversion of the PA effect [10] is as follows,

$$\left(\nabla^2 - \frac{1}{v_s^2}\frac{\partial^2}{\partial t^2}\right) p(\vec{r}, t) = -\frac{\beta}{C_P}\frac{\partial H(\vec{r}, t)}{\partial t} \quad (1)$$

where $v_s$ denotes the speed of sound, $p(\vec{r}, t)$ represents the sound pressure at position $\vec{r}$ and time $t$, $\beta$ represents the thermal expansion coefficient of volume expansion, and $C_P$ represents isobaric specific heat capacity. Further, $H(\vec{r}, t)$ denotes the heat source function, which is defined as the conversion of light energy to heat energy per unit volume and unit time. This formula means that hemoglobin (i.e. blood vessels) specifically absorb the energy of light and radiate sound waves as a photoacoustic transducer.

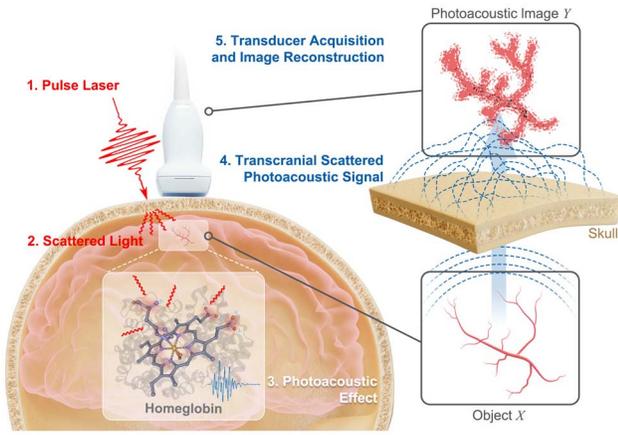

Fig. 1. Schematic of the physical transmission process of PA brain imaging.

In the above process, ultrasonic scattering has a considerably more serious effect on image quality than light scattering because the light offers energy, i.e., information of PA signal amplitude, whereas the ultrasound offers wave information, i.e., information of PA signal amplitude and phase, for PAI. Therefore, for the ultrasonic scattering process, the blood vessel distribution can be considered as the measuring object $X$, the reconstructed PA image from extracranial PA signal can be considered as the measured distribution $Y$, and the process from $X$ to $Y$ as the forward propagation process of the PA signal through the skull. This process can be regarded as a mapping function $F$, and the entire process can be described by a mapping $F : X$ to $Y$. The problem can then be transformed to an optimization problem to be solved to obtain the object information $X$. Thus, we hope to build an appropriate neural network to solve this minimization problem [44] and design this minimization problem into the objective function of the neural network through the design of input and output data and structural risk minimization, to minimize the loss function of the training sample.

$$\hat{X} = argmin \left[ \|F(X) - Y\|_2^2 + \alpha \Phi(X) \right]. \quad (2)$$

### B. Network Architecture

We built a CNN to learn a statistical model relating the speckle patterns and unscattered objects. The overall structure of the proposed CNN (Fig. 2) follows a modified encoder-decoder "U-Net" architecture [45]. Since the U-Net structure was proposed, it has gained widespread attention in the field of medical imaging. The encoder (downsampling)-decoder (upsampling) and skip connection used are a very classic design method. The shallow layer contains high-resolution information of the fine features of the image, and the deep layer contains semantic information of the features between the target and the environment [46]. There are two reasons for using the U-Net structure in photoacoustic imaging through scattering media. First of all, as mentioned above, photoacoustic transcranial imaging is a Mie scattering problem. Therefore, the received transcranial photoacoustic signals used for the reconstruction of cerebrovascular images have the following two characteristics: First, in the frequency domain, the photoacoustic signals generated from the distributed vascular size are broadband. Second, in the spatial domain, the transcranial signal field, which is scattered and superimposed by the size-distributed holes, has a complex spatial distribution. While the shallow and deep layers of U-Net acquire high- and low-frequency information from the image, respectively, and the information of each level is well preserved through skip-connections. Secondly, for transcranial scattering and various medical imaging problems, the imaging target and structure are relatively fixed (compared to natural images). U-Net can reconstruct the spatial structure with self-similarity and use the handcrafted image image and the structure of the known part to interpolate the unknown region [47]. In the results section, we compare U-Net with several other networks.

The network structure is based on U-Net with some modifications. Since blood vessels have distributed sizes, and the U-net structure is little sensitive to very high spatial frequency information [48], there is a deviation in the reconstruction of peripheral blood vessels. So we added the batch normalization (BN) [49] method after each down-sampling convolutional layer, to overcome this problem (as shown in Figure 2). BN effectively normalizes each batch of inputs by re-centering and scaling to alleviate the problem of internal covariate shift. The feature map is normalized as

$$\hat{x}_i^{(k)} = \frac{x_i^{(k)} - \mu_B^{(k)}}{\sqrt{\sigma_B^{(k)^2} + \epsilon}}, \quad (3)$$

where $k$ is the number of convolutional layers, $i$ is the batch size, $\mu_B^{(k)}$ is the mean value of the dimensionality, and $\sigma_B^{(k)^2}$ is the dimensional variance.

In the subsequent step, convolution is performed. The input to the CNN is a preprocessed $128 \times 128$ pixels speckle pattern. Through the "encoder" path, the intermediate output from the encoder has small lateral dimensions but rich depth information. Subsequently, the low-resolution image goes through the "decoder" path, and the information of different spatial scales is transmitted and fused through a skip connection to retain high-frequency information. Finally, a convolutional layer is used to output a specified-size ($128 \times 128$ pixels) image based on the required imaging task.

### C. Simulation Datasets

In the simulation, we used hand-segmented blood vessel images (Fig. 3(a)) from the public Digital Retinal Images for Vessel Extraction (DRIVE) dataset, which contains a total of 40 images. Each image was segmented using a sliding window such that each image had blood vessels with different lengths, widths, tortuosity, branching patterns, and angles. Finally, a total of 4000 blood vessel images were obtained; a representative sample is shown in Fig. 3(b). The open-source k-Wave MATLAB Toolbox [50] was used to simulate the propagation of PA waves derived from blood vessels through porous media. The simulation settings are shown in Fig. 3(c). For the problem of ultrasonic scattering, we simplified the skull and simulated it with a flat porous structure. First, the segmented blood vessel image was binarized and placed in a grid as a PA source with the porous media simulating the skull above it. Then, the PA signal passing through the porous media was received by an ultrasonic linear array probe. Here, we use B-mode for PA brain imaging, which is most convenient for the clinic, instead of C-mode commonly used in the laboratory. Therefore, there is also the problem of incomplete information in the finite angle tomography. The propagation of blood vessel PA signal from 0 $\mu$s to 18 $\mu$s is shown in Fig. 3(d). Finally, the detected PA signals were reconstructed by delay and sum (DAS). In addition, according to the parameters of the skull, the thickness, sound velocity, and density of the porous media (black area) were set to 2.1 mm, 4100 m/s, and 1500 kg/m$^3$, respectively [12], [13]. The rest (the white area, including intracranial pore structures) were set to soft tissues with sound velocity and density of 1500 m/s and 1000 kg/m$^3$, respectively.

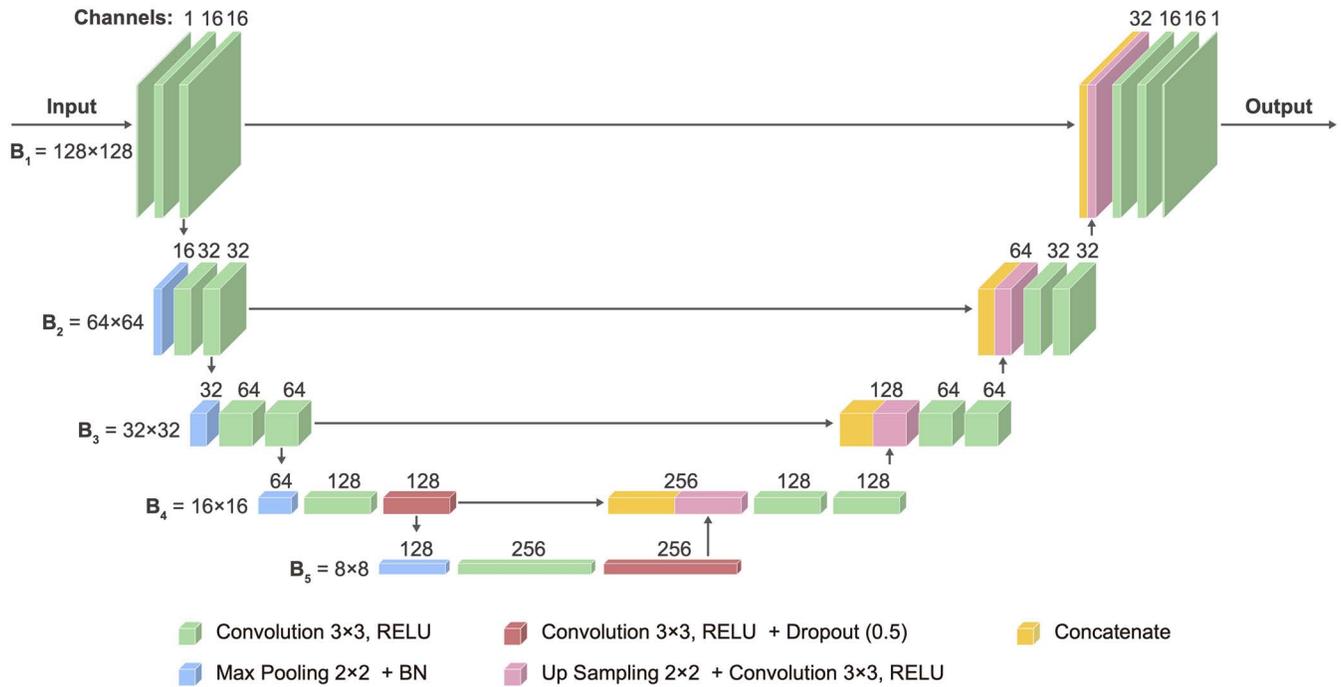

Fig. 2. Proposed convolutional neural network (CNN) model based on U-Net. The number written above each layer indicates the number of convolution kernels (channels). The numbers B1,..., B5 represent the size of the image (the block size in the weight matrix), which remains unchanged in each row.

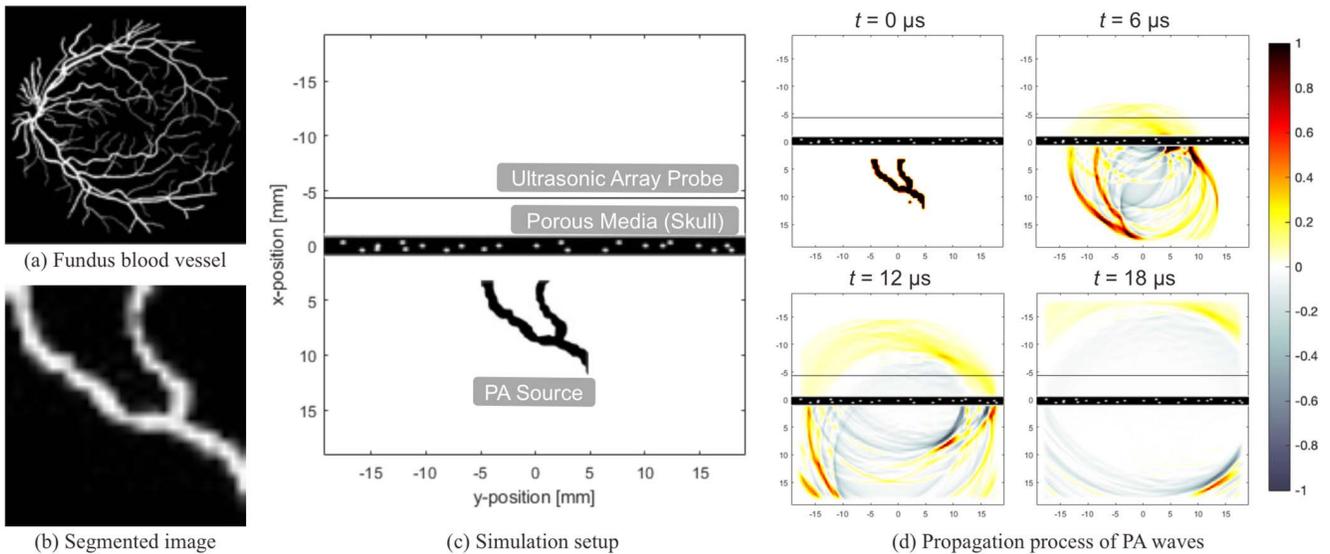

Fig. 3. Simulation of PA signal propagation. (a) Fundus blood vessel image. (b) Segmented blood vessel image from (a) used as the PA source of (c). (c) Simulation setup in k-Wave. (d) Propagation process of PA waves from t = 0 to t = 18 $\mu$s.

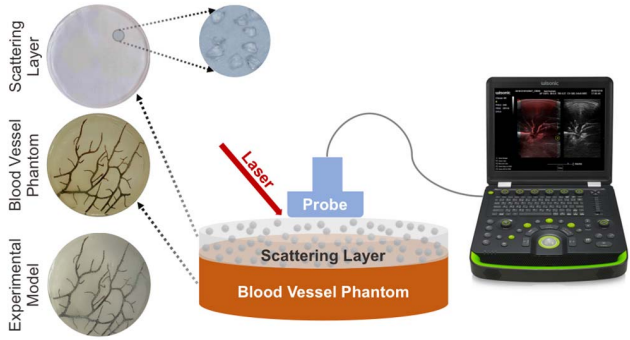

Fig. 4. Experimental setup. On the left side of the figure are the porous media used in the experiment (including the micrograph of the transparent crystal particles), blood vessel phantom, and combined experimental phantom. The positions of the laser and the probe are shown in the middle using a self-developed PAI system (on the right) for acquisition.

The calculation area set in k-Wave is 38.4 mm × 38.4 mm, the time sampling interval is 3 ns, and the total sampling time is 25 $\mu$s. The number of transducer elements is 256, which is defined as a binary linear sensor mask matrix [50], which is used to record the photoacoustic field distribution of the photoacoustic source at each time step. The sensor position is 3 mm above the scattering flat medium.

### D. Phantom Experiment Datasets

The experimental setup is shown in Fig. 4. Coral branches, which were buried in the phantom, (approximately 0.8–1.5 mm in diameter) with random growth directions and bifurcated structures were used to simulate the blood vessel network; they were buried in the phantom. A phantom layer embedded with transparent crystal particles (approximately 100–500 $\mu$m in diameter, corresponding to an ultrasonic frequency of 3–15 MHz) was placed above coral branches to simulate the porous media, and an ultrasonic array probe (selected frequency: 10 MHz) was used on its outside to receive scattered PA signals from the simulated blood vessels. The porous media was then replaced with a homogeneous phantom layer with the same thickness and without crystal particles, and the unscattered PA signal from the simulated blood vessels were received. The instrument for collecting signals and imaging was a self-developed PAI system. The probe frequency is in the range of 6–15 MHz with a center frequency of 10 MHz, ≥90% detection bandwidth. The probe has 128 array elements, and the length, width, and spacing of the array elements are 3.5 mm, 0.2 mm, and 0.2 mm, respectively. The wavelength of the laser used is 750 nm, the pulse frequency of the laser (Phocus Mobile, OPOTEK, Carlsbad, CA) is 10 Hz, and the duration of a single pulse is within 2–5 ns. In preprocessing, we use a 128 × 128 pixels frame to intercept the target position image in the image with/without scattering, and a total of 1000 pairs of images of the imitation blood vessel network with and without scattering were collected.

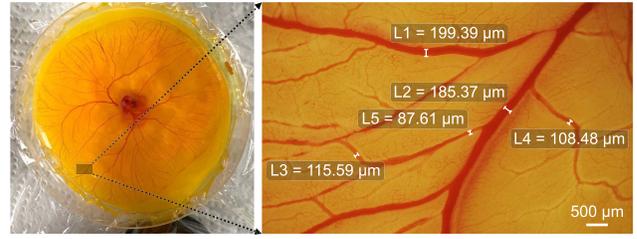

(a) Chick embryo model  (b) Vessel diameter measurement

Fig. 5. Chick embryo experimental model. (a) Representative chick embryo model. (b) Microscopic image with the diameters measurement of the microvessels in the zoom-in area of Fig. 5(a).

### E. Chick Embryo Experiment Datasets

We used a chick embryo model (Fig. 5) to simulate the real blood vessel network *in vivo*. Fig. 5(b) presents the microscopic image and measured sizes of the blood vessels (KH-7700, Hirox, Tokyo, Japan). The fertilized chicken eggs were incubated for 3–5 days, and the blood vessel diameters were approximately 50–300 $\mu$m. The embryos were then removed from the eggshells and placed on a petri dish with a plastic membrane. The experimental system and the porous media were the same as those used in the phantom experiment. Approximately 3–5 sets of images with or without scattering were collected at the same blood vessel position of each egg, a total of 60 eggs. The data were expanded using one-to-one correspondence with or without scattering images. In preprocessing, we use a 256 × 256 pixel frame to intercept the target position image in the image with/without scattering and compress the image with a size of 128 × 128 pixels. Since the chick embryo model has a chick yolk tissue besides blood vessels, we performed noise reduction treatment on it to remove the background noise caused by the chick yolk tissue and obtain a better training effect. We counted the overall gray value of the image and removed the pixel value below 20-% peak gray and later normalize all images. Finally, a total of 778 images were obtained.

### F. Evaluation and Implementation

In this study, we chose the mean absolute error (MAE) as the loss function (see Appendix C for more details.) and Adam as the strategy to update the weights during the training process (learning rate 0.001, dropout rate 0.5, and batch size 64). In the simulation, phantom experiment, and chick embryo experiment, the datasets are divided into a training set, validation set, and test set according to 80%, 10%, and 10%, and training and testing are carried out, respectively. According to the accuracy and loss of the training and validation sets, the epochs of the best models of these three datasets (simulation, phantom experiments, chick embryo experiments) are between 100–150 times, and the total training time is up to 38 h.

The five performance indicators selected for evaluation were MAE, which measures the mean absolute error between the target and speckle image and is used as the loss function and evaluation index simultaneously; mean squared error (MSE), which measures the mean absolute squared between the target and speckle image; structural similarity (SSIM), which measures the similarity of each pair of image brightness, contrast, and structure; multiscale structural similarity index (MS-SSIM), which performs multiscale comprehensive evaluation based on SSIM; and the peak signal-to-noise ratio (PSNR), which objectively indicates image quality.

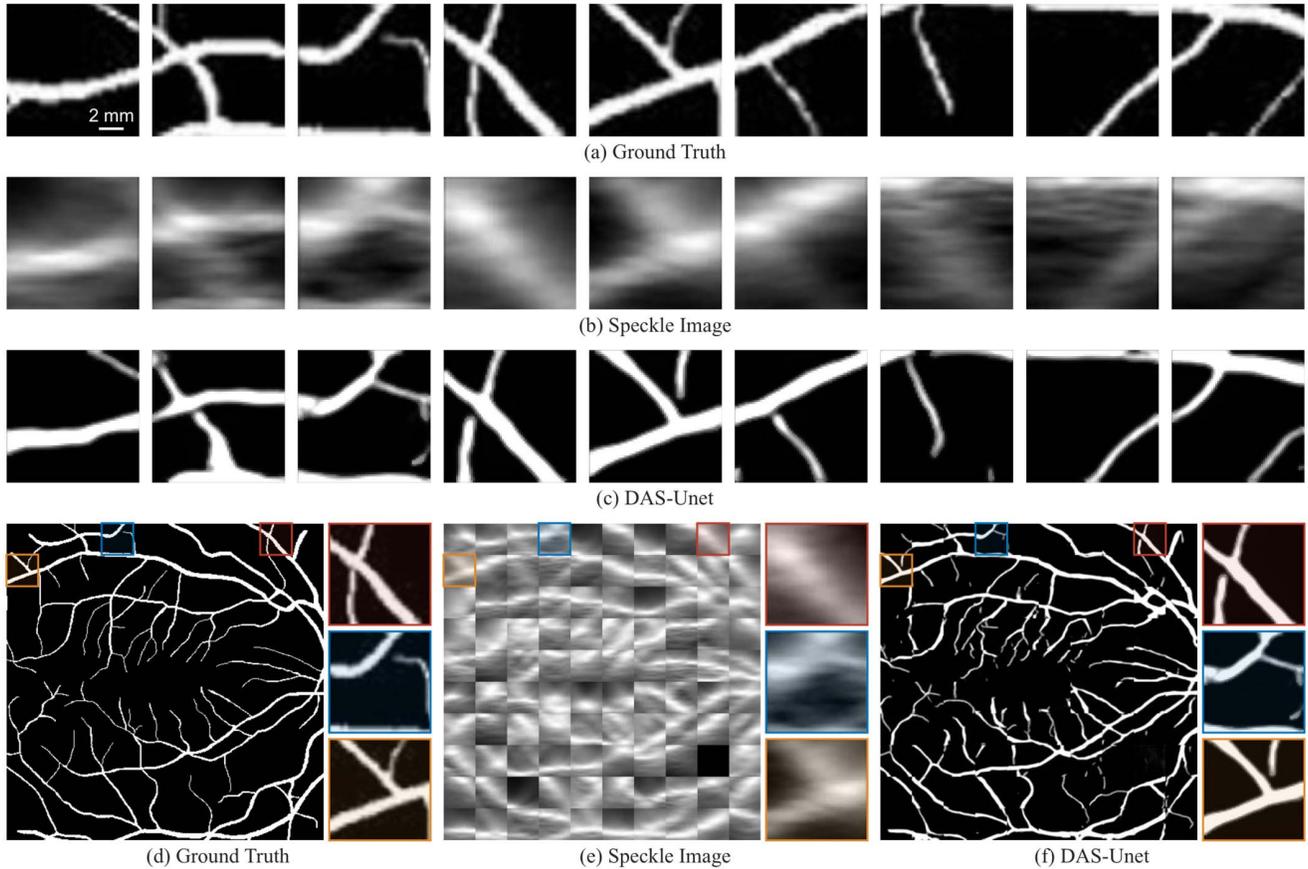

Fig. 6. Reconstructed results of the simulation. (a) The ground truth images in the segmentation database. (b) The speckle images reconstructed by DAS. (c) The reconstructed images using the proposed DAS-Unet. (d–f) the stitched images of (a–c), respectively.

In addition, in image semantic segmentation, especially medical images, Residual Network (ResNet) and Fully Convolutional Network (FCN) are also commonly used network structures for training or performance comparison [51], [52]. Therefore, to verify the applicability of the U-net structure, we built and compared network performance and added ResNet-18 [53] and FCN-8s [54] structures to retrain and predict our data. The epoch selects the best model for storage in the range of 100–150 times according to the performance of the loss function and evaluation indicators. Moreover, we compared this method with some advanced DAS reconstruction algorithms on the simulation data, including delay-multiply-and-sum (DMAS) [55] and multiple-delay-and-sum with enveloping (multi-DASE) [56], see Appendix B for more details.

The CNN was implemented in Keras with the TensorFlow backend (https://keras.io). All computations were conducted on a computer with an Intel Xeon Gold 6130 CPU @ 2.10 GHz, an NVIDIA Quadro P4000 GPU, and 1024 GB of RAM.

## III. RESULTS

### A. Results of Simulation

To test the performance, we randomly selected nine pairs of speckle images from the DRIVE database but not in the training set and input them to the trained CNN. The ground truth images in the segmentation database are shown in Fig. 6(a). The ground truth images are used as a PA source for simulation through porous media, and the speckle images reconstructed by DAS are shown in Fig. 6(b). The reconstructed images from the output of the trained neural network with the images in Fig. 6(a) as input are shown in Fig. 6(c). As it is a post-processing reconstruction method based on DAS, it is called DAS-Unet. Compared with ground truth images, speckle images were accompanied by strong acoustic scattering and acoustic impedance mismatch of the PA waves because of the porous structure of the skull and the difference in acoustic impedance with blood vessel tissues. This can lead to serious speckle artifacts, loss of detail information, and blurred edges; further, it severely degrades the image quality. The CNN output image effectively reconstructed the important features of the target and showed a clear edge and a clean background. In addition, the blood vessel images are obtained by segmentation, and the scattering simulation is performed separately. Therefore, we stitched the images Fig. 6(a-c), as shown in the images Fig. 6(d-f). The results show that a clear and complete image can be obtained very well after stitching.

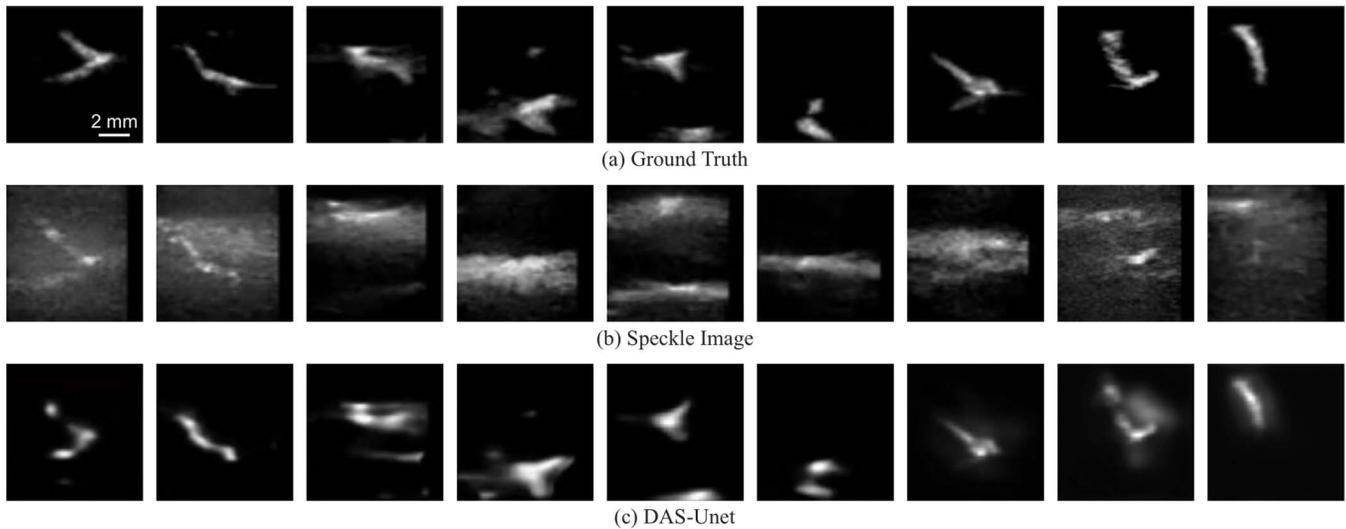

Fig. 7. Reconstructed results of the phantom experiment. (**a**) The ground truth images without porous media. (**b**) The speckle images with porous media. (**c**) The reconstructed images using the proposed DAS-Unet.

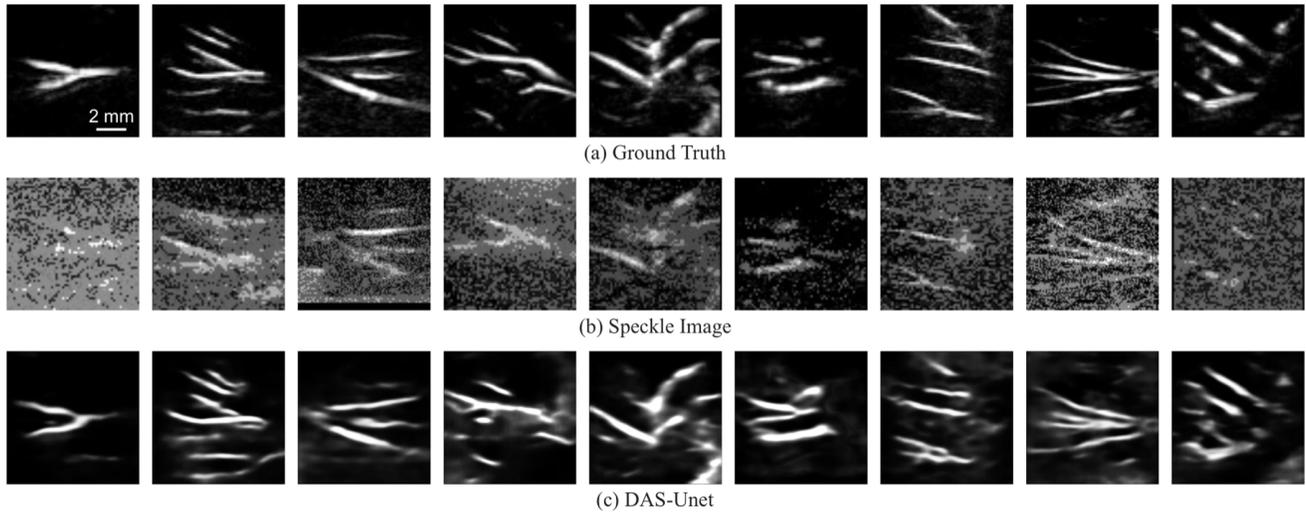

Fig. 8. Reconstructed results of the chick embryo experiment. (**a**) The ground truth images without porous media. (**b**) The speckle images with porous media. (**c**) The reconstructed images using the proposed DAS-Unet.

### B. Results of Phantom Experiment

Similar to the simulation, nine pairs of images were obtained from the experimental data; the images (Fig. 7(b)) collected through the porous media were input to the trained neural network. The output reconstructed image (Fig. 7(c)) was compared with the image collected by the non-porous media (Fig. 7(a)). The images collected by the porous media were accompanied by strong speckle and background noise. After training, the image we obtained restores the structure of the blood vessel network and greatly reduces the background noise.

### C. Results of Chick Embryo Experiment

Images without and with porous media collected from chick embryos are shown in Fig. 8(a) and (b). Compared with the ground truth, the blood vessel image through the porous media is blurred, and the blood vessel information is submerged in noise, making it difficult to accurately identify. Through the training of the DAS-Unet (Fig. 8(c)), the background noise is cleared, and the visibility of the structure is significantly enhanced.

### D. Performance Comparison and Evaluation

We compared the results of DAS-Unet with two other classic network structures commonly used in medical images. The simulation data, phantom experiment, and chicken embryo experiment are input into the network for training and prediction, respectively. The results are shown in Figure 9, which are the ground truth and the reconstruction results of the three methods of simulation, phantom experiment, and chick embryo experiment. Table I lists the MAE, MSE, SSIM, MS-SSIM, and PSNR of all test samples reconstructed the four methods of DAS, DAS-Unet, DAS-ResNet, and DAS-FCN in simulation, phantom experiment, and chick embryo experiment. The method based on DAS-Unet has a better performance than other methods, and the five evaluation indices of MAE, SSIM, and PSNR are greatly improved in the simulation and the experiment. Further, it can quantitatively explain that the blood vessel image reconstructed by DAS-Unet from the speckle image is significantly beneficial in terms of accuracy, brightness, contrast, structural similarity, and image quality.

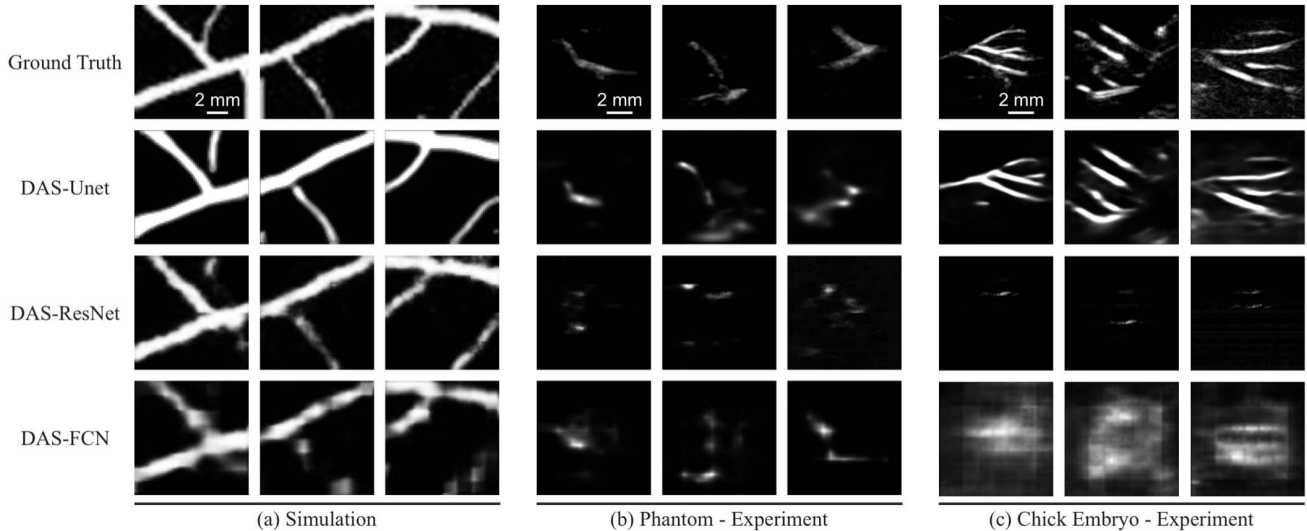

Fig. 9. Comparison of reconstruction results of the (a) simulation, (b) phantom experiment, and (c) chick embryo experiment using DAS-Unet, DAS-ResNet, and DAS-FCN.

TABLE I
PERFORMANCE COMPARISON

| Metrics | Simulation | | | | Phantom - Experiment | | | | Chick Embryo - Experiment | | | |
|---|---|---|---|---|---|---|---|---|---|---|---|---|
| | DAS | DAS-Unet | DAS-ResNet | DAS-FCN | DAS | DAS-Unet | DAS-ResNet | DAS-FCN | DAS | DAS-Unet | DAS-ResNet | DAS-FCN |
| MAE | 0.452 | **0.063** | 0.095 | 0.092 | 0.051 | **0.021** | 0.089 | 0.030 | 0.157 | **0.118** | 0.224 | 0.202 |
| MSE | 0.243 | **0.245** | 0.035 | 0.039 | 0.009 | **0.006** | 0.021 | 0.008 | 0.045 | **0.037** | 0.072 | 0.066 |
| SSIM | 0.046 | **0.670** | 0.424 | 0.547 | 0.394 | **0.788** | 0.094 | 0.630 | 0.073 | **0.214** | 0.040 | 0.080 |
| MS-SSIM | 0.382 | **0.851** | 0.707 | 0.721 | 0.611 | **0.774** | 0.432 | 0.665 | 0.277 | **0.430** | 0.244 | 0.323 |
| PSNR | 6.138 | **16.106** | 14.566 | 14.098 | 20.397 | **22.146** | 16.718 | 21.204 | 13.511 | **14.340** | 11.419 | 11.809 |

## IV. DISCUSSION

To take advantage of PAI, the target media must have relatively uniform acoustic characteristics whereas relatively different light absorption characteristics. However, the skull and brain have significantly different acoustic properties; the heterogeneous porous skull itself, the interface of hard skull and soft brain exhibits a strong acoustic impedance mismatch, which can cause strong scattering and attenuation. Accordingly, time-domain reconstruction algorithms lead to serious speckle noise and scattering artifacts. In addition, there is the incomplete information problem with B-mode imaging as a finite angle tomography technique. Inspired by the performance improvement of CNN in medical imaging, we investigated a method for the PAI of blood vessel networks through thick porous media based on deep learning post-processing. We adopted a combination of CNN and DAS reconstruction algorithms, and captured PA images without/with the porous media as the learning target and the network input, respectively. The complex coding of porous structures is decoded from incomplete information by using CNN feature mining. The results demonstrate that this method successfully removes speckles and artifacts and outputs a PA image close to the real blood vessel network.

This method is mainly based on data-driven learning, which means that the quality of the final reconstructed image largely depends on the quality of the training dataset. In the simulation, we used the segmented blood vessel image as the ground truth. The parameter setting was ideal, and the influence of the system and environmental noise was neglected. Therefore, it was straightforward to fit the mapping relationship of the imaging system following training. However, in the experiment, the PA image collected by the non-porous media was used as the ground truth, which already contained some background noise. In the coral branch phantom experiment, as the imaged structure is a three-dimensional structure, its cross-sectional imaging can only obtain simple structures with similar points or Y-shaped bifurcations; hence, the reconstruction image has a high evaluation index. In the chick embryo imaging experiment, the blood vessels are distributed on a relatively flat plane; thus, images with rich blood vessel information could be obtained. However, compared with the real image, the MAE and SSIM would be lower, and the background noise would also result in a higher PSNR. Zhang *et al.* proposed a power azimuth spectrum method [3], overcoming the low signal-to-noise ratio in PAI and showed the size and structural direction of microvessels visually. In future research, the power azimuth spectrum can be combined with our method to image vascular structure and fully extract vascular parameters to further evaluate the vascular network.

However, this deep learning-based method invariably has some limitations, such as low robustness, high cost, and time-consuming data acquisition and network training in the early stage. In this study, we obtained datasets based on a specific porous structure and trained the network. Whereas, a well-trained network can only be used to predict the imaging of this specific structure, which greatly limits the application of this method. Nevertheless, by learning the statistical information contained in the speckle intensity patterns captured on a set of diffusers with the same macroscopic parameter, the trained CNN can generalize and produce high-quality object predictions through an entirely different set of diffusers of the same class [42]. For clinical application, each skull has a different but similar structural distribution. This method may overcome this limitation; through the learning of multiple isolated skulls, a CNN can ultimately learn the general porous characteristics of the skull to reconstruct high-quality transcranial PA images.

The most important limitation is that the current deep learning method cannot reveal the physical mechanism of sound propagation in porous media, although it can determine the mapping relationship between the input and output of an imaging system. Many researchers have noted this problem and have attempted to integrate physical relationships [57] and mathematical methods [58], [59] with deep learning. Specifically, they used the physical relationship to constrain and correct the learned mapping relationship, so that it may better fit physical principles. This will assist deep learning in achieving critical breakthroughs and progress in the field of imaging.

In this study, we consider the mixing problem of a broadband source being complexly scattered and superimposed through an irregular pore medium. Therefore, we believe that the method may be effective in solving the decoding problem of propagation of more broadband signal sources through complex paths. The high sensitivity of photoacoustic imaging to endogenous chromophores [60], [61] and exogenous probes [9], [62] of the brain gives this modality an unprecedented ability to interrogate the brain non-invasively in physiological and disease conditions [63]. For instance, Deliolanis *et al.* demonstrated the ability of photoacoustic imaging to visualize deep-seated glioma brain tumors for intraoperative depiction of tumor margins [64]. Lv *et al.* used photoacoustic imaging to visualize the localization of infarct site and monitor blood oxygen function in an ischemic stroke model [65]. These application have yielded promising results in mouse models. However, human brain imaging facing clinical translation still needs to overcome the thick cranial bone so that photoacoustic signals containing rich tissue information can pass through the skull and be received and decoded. In addition, the same problem of complex path propagation exists for bone imaging which is an important evaluation modality for orthopedic diseases such as osteoporosis. In photoacoustic imaging, porous bone tissue is both a photoacoustic source and a strong scattering medium [66], and tiny bone trabeculae and bone marrow are excited to broadband photoacoustic signals at different optical wavelength [67]. However, when their photoacoustic signals passing through porous bone tissue, they are accompanied by the complex scattering superimposed effects of broadband signals in frequency-domain and spatial-domain.

Next, we hope to continue to extend the method to transcranial imaging, in addition to the aforementioned problem of individual differences in the skull, another very important challenge is its irregular three-layer structure of cortical bone, cancellous bone, and cortical bone [12], [14]. There is strong shear wave propagation in cortical bone and complex wave mode conversion at the boundary between cancellous and dense bone [68]. The wave vector decomposition and transfer matrix methods will be an effective approach to solve this problem.

## V. CONCLUSION

In this study, we proposed a CNN based on U-Net to learn the mapping relationship between the speckle pattern and the target for finite-angle PAI under thick porous media. The results demonstrate the superiority of the proposed method to extract effective information from highly blurred speckle patterns to reconstruct the target image quickly. The proposed network provides an effective tool for the image quality improvement of transcranial PA brain imaging.

As is inherent in all learning approaches, the limitations of the proposed method are dictated by the training datasets. In future research, we will consider combining the CNN with a physical model and study a series of PAI experiments involving intracranial blood vessels in rabbits and humans to verify the effectiveness of deep learning methods in scattering decoding.

## APPENDIX A
## SCATTERING MODEL

The type of scattering in a transcranial imaging model depends on the size of the skull pores and the wavelength of the acoustic signal. The skull is composed of three layers: cortical bone, cancellous bone, and cortical bone [12], [14]. The pore sizes of the human cortical and cancellous bones are approximately within the range of 10 $\mu$m–2 mm [12], [14]–[19], which is a distributed pore size.

Regarding the scattering model, there are some differences between photoacoustic imaging and traditional ultrasound imaging (Fig. 10) as follows.

Ultrasound imaging: The probe excites single-frequency ultrasound and irradiates it into the tissue. Owing to the difference in the acoustic impedance of each layer of biological tissues, the reflected wave returns to the probe to form an acoustic signal. Then, the acoustic signal reflected from the brain tissue received by the probe passes through the skull twice. In the process of traversing, due to the distribution of skull pore size, there is a high scattering loss in a wide frequency band, especially high frequency [69], [70]. Therefore, low-frequency incident sound waves (E.g., 1.3 MHz) with less energy loss have to be used to maintain a sufficient signal-to-noise ratio. At this time, the scattering of low-frequency sound waves ($\lambda = 2- 3$ mm@2500–4100 m/s) in the skull by the skull pores (0.3 mm) is mainly Rayleigh scattering [71] as shown in Fig. 11, and the imaging resolution is also low.

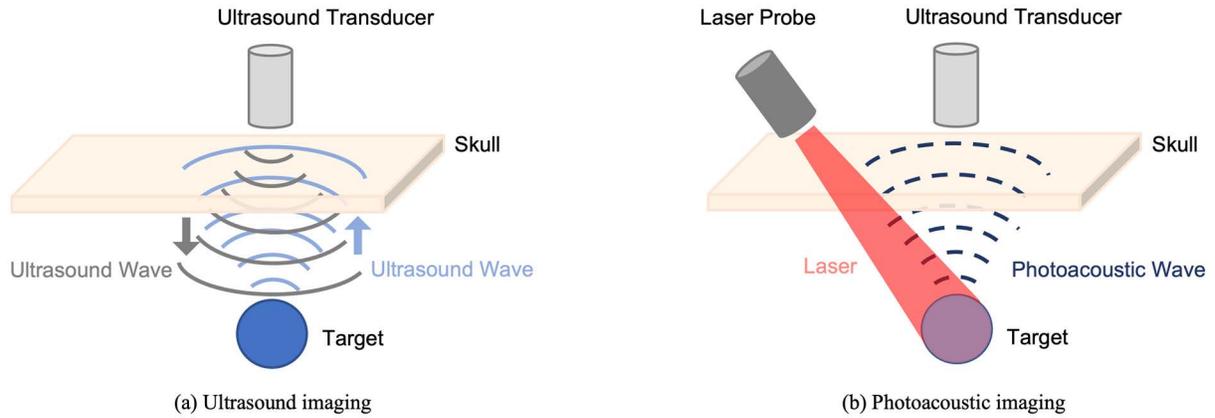

Fig. 10. Acoustic signal generation in (a) ultrasound imaging and (b) photoacoustic imaging.

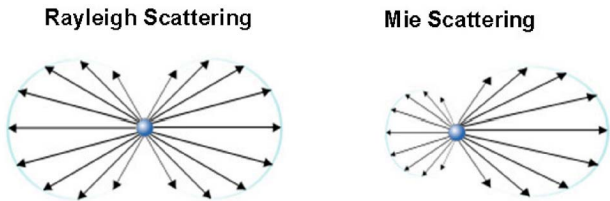

Fig. 11. Comparison of scattering intensity of Rayleigh scattering and Mie scattering at various angles [73].

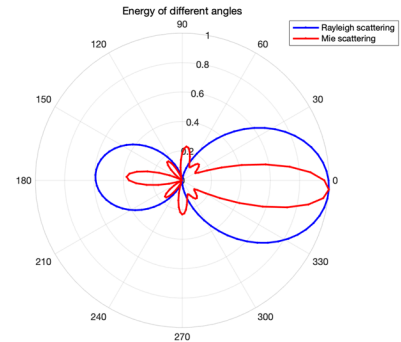

Fig. 12. Comparison of scattering intensity at various angles based on the parameters in Table II.

Photoacoustic imaging [10], [72]: Firstly, the pulsed laser excites the vibrational energy level of the intracranial chromophore (such as hemoglobin) to produce ultrasonic waves, which is called the photoacoustic effect, and then, the probe receives these ultrasonic waves that pass through the skull. Because these ultrasonic waves only pass through the skull once, their transcranial scattering attenuation is much less than that of ultrasound imaging, which allows more high-frequency ultrasonic waves to penetrate the skull and be received by transducers. Secondly, the frequency of the photoacoustic signal is determined by the size of the chromophore and the speed of sound of the medium under ultra-short pulse laser excitation (2–5 ns in this work). In this work, we focused on cerebral microvessels with a diameter of 100–200 $\mu$m, so the excited photoacoustic signal has a wide bandwidth of 7.5–15 MHz with the sound speed of soft tissue of 1500 m/s [10], [11]. When transmitted to the skull, their wavelengths are approximately in the range of 167–547 $\mu$m according to the sound speed of the skull (2500–4100 m/s), which is of the same order as the size of a skull pore (100–500 $\mu$m in this work), so it is Mie scattering here [20], [21]. Fortunately, in Mie scattering, the scattered energy is concentrated in the forward direction benefiting photoacoustic imaging, as shown in Fig. 12. Unfortunately, unlike Rayleigh scattering, Mie scattering is accompanied by some side lobes on both sides of the main lobe, so the beamforming process is more complicated. Therefore, transcranial photoacoustic microvascular imaging combines the broadband scattering problem in the frequency domain with the main lobe and side lobes signal superimposition problems in the spatial domain. That is why we attempted to use the U-Net structure to obtain wide-band scattering feature recognition, which can strengthen the correlation of spatial information at different scales and restore satisfactory high-resolution vascular photoacoustic imaging.

## APPENDIX B
## COMPARISON OF RECONSTRUCTION ALGORITHMS

We choose DAS as the original data reconstruction algorithm because this method can retain the original and rich information of the signal to a large extent, which is convenient for the neural network to mine and extract effective information. In addition, we compared this method with some advanced DAS reconstruction algorithms on the simulation data, including DMAS [55] and multi-DASE [56]. Fig. 13 shows that these optimization algorithms effectively reduce image noise, but unfortunately, these are not suitable for scattering media. The image artifacts are still strong, and while noise is removed, some tiny blood vessel information is also weakened.

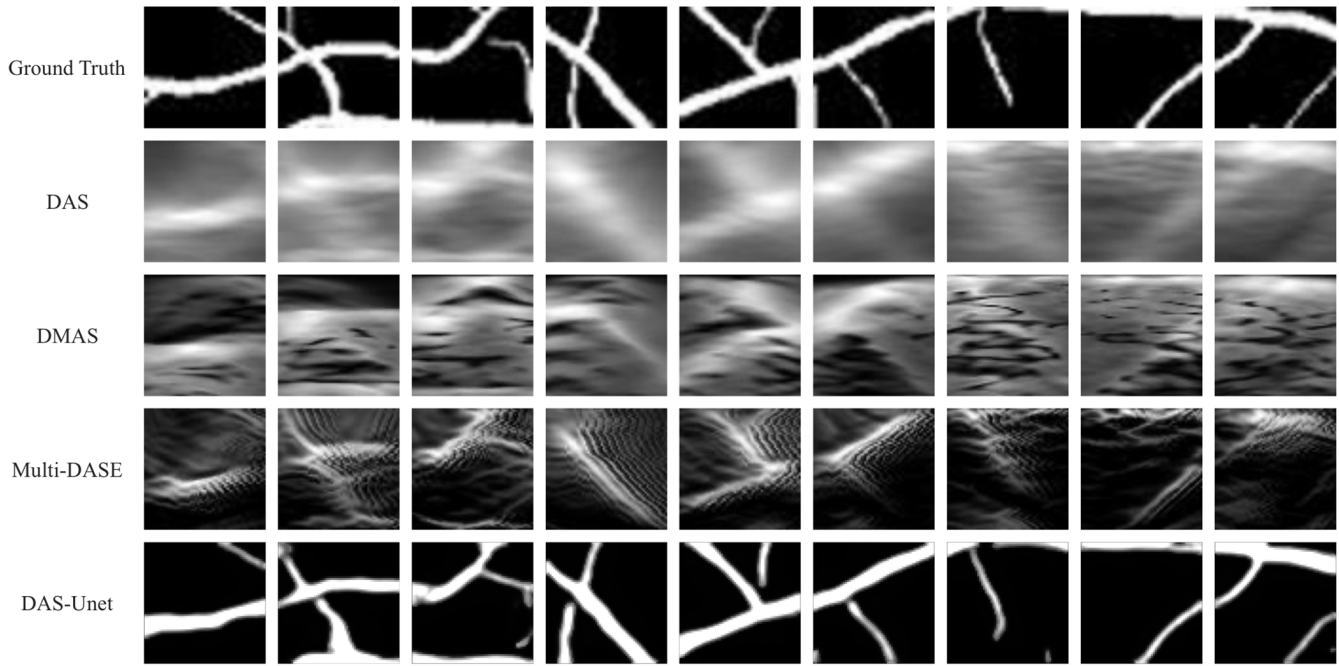

Fig. 13. Reconstruction algorithm comparison.

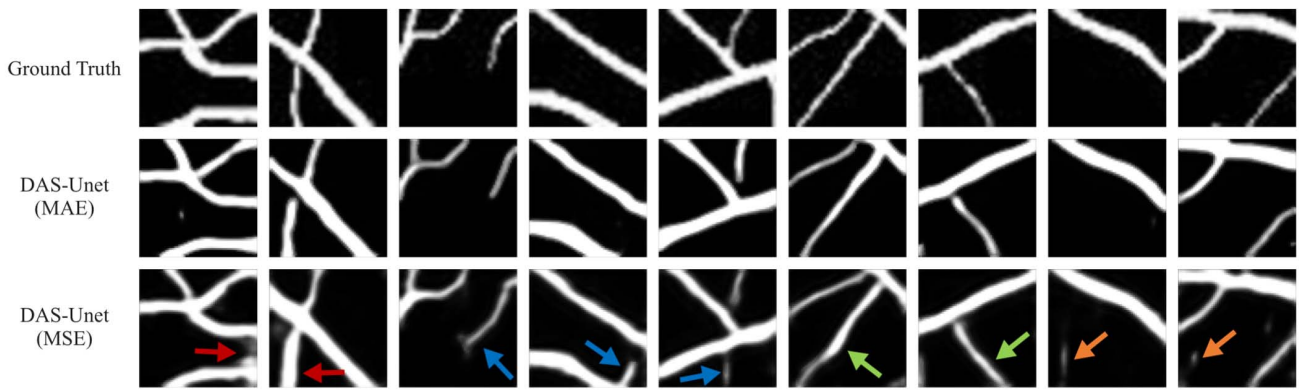

Fig. 14. Comparison of simulation results using loss functions MAE and MSE.

TABLE II
COMPARISON OF PHOTOACOUSTIC/ULTRASOUND TRANSCRANIAL IMAGING

|  |  | Ultrasound imaging | Photoacoustics imaging |
|---|---|---|---|
| Scattering mode |  | Rayleigh scattering | Mie scattering |
| Transcranial times |  | 2 | 1 |
| Ultrasound frequency |  | Narrow band | Wide band |
| Forward scattering distribution field |  | Single main lobe | Main and side lobes |
| Scattering parameters (example) | Hole size | 300 μm | 300 μm |
|  | Wavelength | 2500 μm | 350 μm |

APPENDIX C
LOSS FUNCTION

Regarding the choice of the loss function, we tested these two loss functions during the training process, and the simulation results using MAE and MSE, respectively, are shown in Fig.14. We found that using MSE can easily result in thickening of blood vessels (red arrow), lengthening (blue arrow), distortion of the bending direction (green arrow), and false noise (orange arrow) in the avascular area. We think that this might result from the image differences. For example, the simulated images are directly separated, and there is little blood vessel information in some images. These images are treated as anomalous images and given a lot of weight when using MSE training. This causes the training to be updated in reducing the error of abnormal points at the expense of other samples, which reduces the overall performance of the model. Conversely, MAE treats abnormal points as damaged data, and the gradient update is more stable. So, it is more suitable for our data. Therefore, we chose MAE as the loss function.

TABLE III
PERFORMANCE COMPARISON BY USING LOSS FUNCTIONS MAE AND MSE

| Metrics | Simulation | | |
|---|---|---|---|
| | DAS | DAS-Unet (MAE) | DAS-Unet (MSE) |
| MAE | 0.452 | 0.063 | 0.685 |
| MSE | 0.243 | 0.245 | 0.230 |
| SSIM | 0.046 | 0.670 | 0.580 |
| MS-SSIM | 0.382 | 0.851 | 0.821 |
| PSNR | 6.138 | 16.106 | 16.393 |

TABLE IV
COMPARISON OF NON-REFERENCE EVALUATION PARAMETERS BETWEEN GROUND TRUTH AND DAS-UNET

| Metrics | Ground Truth | DAS-Unet |
|---|---|---|
| Laplacian | 0.010 | 0.026 |
| Variance | 300.418 | 277.505 |

## APPENDIX D
## NON-REFERENCE EVALUATION

All the above evaluation indicators are based on ground truth for reference evaluation, and we added two evaluation parameters to evaluate the two sets of images of the ground truth and DAS-Unet without reference. The Laplacian value and variance value of the image.

$$\text{Laplacian}: \quad D(f) = \sum_y \sum_x |G(x, y)|,$$

where $G(x, y)$ is the convolution of the Laplacian operator at the pixel point $(x, y)$. This function evaluates the sharpness of the image by calculating the image gradient value.

$$\text{Variance}: \quad D(f) = \sum_y \sum_x (|f(x, y) - \mu|,$$

where $\mu$ represents the average gray value of the entire image. This function is more sensitive to noise. The purer the image, the smaller the value of the function.

It can be seen from the Table IV that the definition of the predicted image we finally obtained is even higher than the ground truth as the learning target, which shows that this method has great potential to improve the quality of the photoacoustic image of a scattered medium.

## ACKNOWLEDGMENT

The authors would like to thank Editage (www.editage.cn) for English language editing.